\documentclass[pra,showpacs,showkeys,preprintnumbers,amsmath,amssymb]{revtex4}
\usepackage{tipa}
\usepackage{amssymb}
\usepackage{txfonts}
\usepackage{mathrsfs}
\usepackage{hyperref}
\usepackage{graphicx}
\usepackage{epsfig}

\begin{document}
\title{A Novel Model of Charged Leptons$^{*}$}
\author{Dianfu Wang$^{\dag}$, Xiao Liang, Yanqing Guo$^{\ddag}$\footnotetext{ $^{\dag}$ Email: wangdfu@dlmu.edu.cn.\\ $^{\ddag}$
Email: yqguo@dlmu.edu.cn.}} \affiliation{School of Science, Dalian
Maritime University, Dalian, China} 

\begin{abstract}
A novel model of charged leptons is presented, which contains two
basics hypotheses. The first hypothesis is that the Yukawa
coupling between Higgs field and charged leptons is the weak
interaction, the Higgs field is a scalar intermediate boson which
changes the chirality of charged leptons in the weak interaction.
The other hypothesis is that the flavor eigenstates of charged
leptons are the superposition states of left-handed and
right-handed elementary Weyl spinors before the electroweak
symmetry breaking. According to this model, the Yukawa coupling
constants between Higgs field and three generations of charged
leptons are considered to be a universal constant, and the
difference of the masses of different charged leptons is due to
the different left-right mixing angles of their flavor
eigenstates.
\end{abstract}
\maketitle


\section{Introduction}

Up to now, with the discovery of the 125 GeV Higgs boson in 2012
at the CERN LHC [1, 2], the  Weinberg-Salam (WS) model [3, 4] of
the electroweak interaction of particle physics stands triumphant,
and almost all relevant experimental results in particle physics
are consistent with this model. Although the WS model has achieved
impressive success in correlating all observed low-energy data in
terms of a very few parameters, it cannot be called perfect. For
example, the model is based on too many assumptions and leaves
many fundamental questions unanswered. The success of the WS model
only involves the gauge sector of the theory, in which only one
free parameter, the Weinberg angle $\theta_{W}$, is used to
understand numerous neutral-current data. But for the fermionic
sector, the fermion mass spectrum ranges from $170 GeV$ of the
top-quark to $0.511\times10^{-3} GeV$ of the electron. We do not
know why there exists such a large difference among the masses of
these fermions, and we have no deep understanding of the origin of
these masses. Since the fermion masses are related to the Yukawa
couplings, we can only understand these differences and the
mass-generation phenomenon by studying the Yukawa couplings in
detail [5, 6, 7, 8].

\section{A Charged Lepton Model}

Quarks and leptons constitute the basic building blocks of matter
in the standard model. In the WS model, there are three
generations of quarks and leptons with identical quantum numbers
but different masses. Since quarks are supposed to be confined by
virtue of their strong interactions, the meaning of the quark mass
is not obvious [9]. Therefore, in the present paper, for
simplicity, we will limit ourselves to three generations of
leptons. Let's start with discussing the lepton sector.
Observationally, we must incorporate a neutral, left-handed
neutrino (For simplicity, the neutrino is assumed to be massless,
and hence right-handed neutrino is absent.) along with a charged
lepton, which can be considered to be the sum of left-handed and
right-handed Weyl spinors. The left-handed fermions form an
isodoublet, consisting of the neutrino and the charged lepton:

\begin{equation}
L_{\ell}=
\frac{1}{2}(1+\gamma_{5})\left(\begin{matrix}\nu_{\ell}\\\ell\end{matrix}\right)=\left(\begin{matrix}\nu_{\ell}\\\ell\end{matrix}\right)_{L},
\end{equation}
while the right-handed sector consists of an isosinglet, the
right-handed charged lepton:
\begin{equation}
R_{\ell}=\frac{1}{2}(1-\gamma_{5})\ell=\ell_{R},
\end{equation}
where $\ell=e,\mu,\tau$ denote the flavor indexes of the charged
leptons, and the subscripts $L$ and $R$ refer to the left- and
right-handed components, respectively. In the standard WS model,
the usual default setting is that the flavor eigenstates of
charged leptons are the same as their mass eigenstates ( An
eigenstate of finite mass is a superposition of left and
right-handed states with equal weight [9]. ), i.e.,
\begin{equation}
\ell=\ell_{L}+\ell_{R}.
\end{equation}

With the chiral fermions given in Eq.(1) and Eq.(2), the most
general $SU(2)_{L}\times U(1)_{Y}$ gauge-invariant lepton part of
the Lagrangian density of WS model
 can be written as

\begin{flalign}
\begin{split}
\mathscr{L}_{\ell}=&-\overline{L}_{\ell}\gamma_{\mu}\left(\partial_{\mu}-ig\frac{1}{2}\tau^{i}A_{\mu}^{i}+ig'\frac{1}{2}B_{\mu}\right)L_{\ell}\\
&-\overline{R}_{\ell}\gamma_{\mu}\left(\partial_{\mu}+ig'B_{\mu}\right)R_{\ell}-G_{\ell}\left(\overline{L}_{\ell}\phi
R_{\ell}+\overline{R}_{\ell}\phi^{\dag} L_{\ell}\right).
\end{split}
\end{flalign}
By defining the mass eigenstates of gauge fields

\begin{flalign}
\begin{split}
&W_{\mu}^{\pm}=\frac{1}{\sqrt{2}}\left(A_{\mu}^{1}\mp iA_{\mu}^{2}\right),\\
&Z_{\mu}=\sin\theta_{W}B_{\mu}-\cos\theta_{W}A_{\mu}^{3},\\
&A_{\mu}=\cos\theta_{W}B_{\mu}+\sin\theta_{W}A_{\mu}^{3},
\end{split}
\end{flalign}
where $\theta_{W}$ is the Weinberg angle with
$\tan\theta_{W}=g'/g$, the interaction term that leptons coupled
to gauge fields and Higgs field in Eq.(4) can be written as

\begin{flalign}
\begin{split}
\mathscr{L}_{\ell}'=&+i\frac{g}{\sqrt{2}}\left[W_{\mu}^{+}\overline{({\nu}_{\ell})}_{L}\gamma_{\mu}\ell_{L}+W_{\mu}^{-}\overline{\ell}_{L}\gamma_{\mu}{({\nu}_{\ell})}_{L}\right]\\
&-i\sqrt{g^{2}+g'^{2}}Z_{\mu}\left[\frac{\overline{({\nu}_{\ell})}_{L}\gamma_{\mu}({\nu}_{\ell})_{L}-\overline{\ell}_{L}\gamma_{\mu}\ell_{L}}{2}+\sin^{2}\theta_{W}\overline{\ell}\gamma_{\mu}\ell\right]\\
&-i\frac{gg'}{\sqrt{g^{2}+g'^{2}}}A_{\mu}\overline{\ell}\gamma_{\mu}\ell
-G_{\ell}\left(\overline{L}_{\ell}\phi
R_{\ell}+\overline{R}_{\ell}\phi^{\dag} L_{\ell}\right).
\end{split}
\end{flalign}
The last term on the right side of Eq.(6) is the gauge invariant
Yukawa coupling between Higgs field and fermions,
\begin{equation}
\mathscr{L}_{Y}=-G_{\ell}\left(\overline{L}_{\ell}\phi
R_{\ell}+\overline{R}_{\ell}\phi^{\dag} L_{\ell}\right),
\end{equation}
where $G_{\ell}$, an additional parameter, gives the strength of
the Yukawa coupling. In the standard WS model, all the charged
leptons are massless as long as electroweak symmetry is unbroken,
and they can get masses from Yukawa interactions only after
electroweak symmetry breaking. A convenient way to implement this
is to introduce a doublet Higgs field
\begin{equation}
\phi=\left(\begin{matrix}\phi_{+}\\\phi_{0}\end{matrix}\right),
\end{equation}
where the subscripts refer to the electric charge. Replacing
$\phi$ by its vacuum expectation value
\begin{equation}
\langle\phi\rangle=\left(\begin{matrix}0\\\upsilon\end{matrix}\right),
\end{equation}
the mass term of the charged leptons can be given by the Yukawa
coupling Eq.(7) as
\begin{equation}
\mathscr{L}_{m_{\ell}}=-{G_{\ell}\upsilon}\overline{\ell}\ell.
\end{equation}
Thus we can identify the masses of the charged leptons
\begin{equation}
m_{\ell}={G_{\ell}\upsilon}.
\end{equation}
However, it does not specify the value of the masses since the
Yukawa coupling constant $G_{\ell}$ is arbitrary. So far, the
standard model of the electroweak interaction has not given more
further information on the origin of charged lepton masses.
Furthermore, it is important to notice that since the strength of
interaction between the Higgs field and any particle is
proportional to the mass of that particle, this means that there
is a new type of interaction that is different from the strong
interaction, the electroweak interaction and even the
gravitational interaction, which is very unnatural.

Since the mass of the Higgs boson is similar to that of W and Z
bosons, so we could find it on the LHC in 2012. To some extent,
they are all the same particles, or fields that presumably mediate
weak interactions. In this sense, the Higgs boson will be no
different from W and Z bosons. If these considerations are
reasonable, the  coupling constants of the four interaction terms
in Eq.(6) will be of the same order of magnitude, which means the
Yukawa coupling in Eq.(4) can be written as
\begin{equation}
\mathscr{L}_{Y}=-g''\left(\overline{L}_{\ell}\phi
R_{\ell}+\overline{R}_{\ell}\phi^{\dag} L_{\ell}\right),
\end{equation}
where $g''$ is the universal coupling constant of the Yukawa
coupling. Based on the above discussion, we assume that the Yukawa
coupling is the weak interaction, and the Higgs field is a scalar
intermediate boson which changes the chirality of particles in the
weak interaction. Since in low-energy processes the Higgs-meson
and the W-meson carry small momentum, the propagators of them may
be taken to be $M_{H}^{-2}$ and $M_{W}^{-2}$. By considering
Eq.(6) and Eq.(12), we can then obtain the following relation
between the coupling constant $g''$ and the gauge coupling
constant $g$ as
\begin{equation}
\frac{g^{2}}{2M_{W}^{2}}=\frac{g''^{2}}{M_{H}^{2}},
\end{equation}
which gives the universal Yukawa coupling constant
\begin{equation}
g''=\frac{g}{\sqrt{2}}\frac{M_{H}}{M_{W}}
\end{equation}

Replacing $\phi$ by its vacuum expectation value
$\langle\phi\rangle$, Eq.(12) changes to be the mass term of the
charged leptons
\begin{equation}
\mathscr{L}_{m_{\ell}}=-{g''\upsilon}\overline{\ell}\ell=-\frac{g\upsilon}{\sqrt{2}}\frac{M_{H}}{M_{W}}\overline{\ell}\ell,
\end{equation}
The massess of the charged leptons derived from Eq.(15) can be
easily shown as
\begin{equation}
m_{\ell}=\frac{g\upsilon}{\sqrt{2}}\frac{M_{H}}{M_{W}}=\sqrt{2}{M_{H}}.
\end{equation}
The comparison between Eq.(16) and Eq.(11) shows that this result
is obviously wrong. The question now is whether the assumption
Eq.(12) is wrong or our understanding of it is incomplete. If we
accept Eq.(12), the only thing we can do is to reinterpret the
flavor eigenstates of the charged leptons.

The form of the Lagrangian density given by Eq.(4) indicates that
the basic spinor fermion fields are not 4-component Dirac spinors,
but rather their left and right-handed projections. Thus, for the
weak interactions, the elementary entities are states of chirality
with zero mass. Based on these viewpoints, we propose a hypothesis
here that the flavor eigenstates of the charged leptons are the
superposition states of left-handed and right-handed Weyl spinors
before electroweak symmetry breaking, i.e.,
\begin{equation}
\psi_{\ell}=\sqrt{2}\left(\cos\theta_{\ell}\ell_{L}+\sin\theta_{\ell}\ell_{R}\right).
\end{equation}
Where $\theta_{\ell}$ are three left-right mixing angles (
Hereafter, we will limit the angles $\theta_{\ell}$ to
$0\leq\theta_{\ell}\leq\pi/2$. ), and factor $\sqrt{2}$ is to
ensure that Eq.(17) is the same as Eq.(3) when
$\cos\theta_{\ell}=\sin\theta_{\ell}=1/\sqrt{2}$. $\ell_{L}$ and
$\ell_{R}$ refer to the elementary entities, the left- and
right-handed Weyl spinors, respectively.

By using Eq.(17), the left- and right-handed charged leptons given
in Eq.(1) and Eq.(2) should be replaced by
\begin{flalign}
\begin{split}
&L_{\ell}'=\frac{1}{2}(1+\gamma_{5})\left(\begin{matrix}\nu_{\ell}\\\psi_{\ell}\end{matrix}\right)=\left(\begin{matrix}\nu_{\ell}\\\psi_{\ell}\end{matrix}\right)_{L},\\
&R_{\ell}'=\frac{1}{2}(1-\gamma_{5})\psi_{\ell}=(\psi_{\ell})_{R}.
\end{split}
\end{flalign}
Consequently, the Yukawa coupling in Eq.(4) becomes
\begin{equation}
\mathscr{L}_{Y}=-\frac{g}{\sqrt{2}}\frac{M_{H}}{M_{W}}\left(\overline{{L}_{\ell}'}\phi
R_{\ell}'+\overline{{R}_{\ell}'}\phi^{\dag} L_{\ell}'\right).
\end{equation}
By using Eq.(18) and Eq.(19), the Lagrangian density which
characterizes the coupling of leptons to gauge fields and Higgs
field in Eq.(4) changes to be
\begin{flalign}
\begin{split}
\mathscr{L}_{\ell}'=&+i\frac{g}{\sqrt{2}}\left[W_{\mu}^{+}\overline{({\nu}_{\ell})}_{L}\gamma_{\mu}{(\psi_{\ell})}_{L}+W_{\mu}^{-}\overline{(\psi_{\ell})}_{L}\gamma_{\mu}{({\nu}_{\ell})}_{L}\right]\\
&-\frac{i}{2}\sqrt{g^{2}+g'^{2}}Z_{\mu}\left[{\overline{({\nu}_{\ell})}_{L}\gamma_{\mu}({\nu}_{\ell})_{L}
-\overline{(\psi_{\ell})}_{L}\gamma_{\mu}(\psi_{\ell})_{L}}\right. \\
&\left.+2\sin^{2}\theta_{W}\overline{\psi}_{\ell}\gamma_{\mu}{\psi}_{\ell}\right]-i\frac{gg'}{\sqrt{g^{2}+g'^{2}}}A_{\mu}\overline{\psi}_{\ell}\gamma_{\mu}{\psi}_{\ell}\\
&-\frac{g}{\sqrt{2}}\frac{M_{H}}{M_{W}}\left(\overline{{L}_{\ell}'}\phi
R_{\ell}'+\overline{{R}_{\ell}'}\phi^{\dag} L_{\ell}'\right).
\end{split}
\end{flalign}
In Eq.(20), all the terms except the Yukawa term do not involve
cross term of left- and right-handed states. Thus, in these terms,
according to quantum principles, it can be seen from Eq.(17) that
the left-handed fermion state
$(\psi_{\ell})_{L}=\sqrt{2}cos\theta_{\ell}\ell_{L}$ and
$\ell_{L}$ correspond to the same state, and for the same reason,
the right-handed fermion state
$(\psi_{\ell})_{R}=\sqrt{2}sin\theta_{\ell}\ell_{R}$ and
$\ell_{R}$ correspond to the same state, too. However, in the
Yukawa term, because of the interference between left-handed state
$(\psi_{\ell})_{L}$ and right-handed state $(\psi_{\ell})_{R}$,
the coefficients $\sqrt{2}cos\theta_{\ell}$ and
$\sqrt{2}sin\theta_{\ell}$ are indispensable.

Based on the above considerations, after parameterization and
unitary gauge transformation, replacing $\phi$ by
\begin{equation}
\left(\begin{matrix}0\\\upsilon+\eta\end{matrix}\right),
\end{equation}
where $\eta$ is the real Higgs field. The Yukawa term in Eq.(20)
changes to be
\begin{equation}
\mathscr{L}_{Y}=-{\frac{g}{\sqrt{2}}\frac{M_{H}}{M_{W}}\sin(2\theta_{\ell})(\upsilon+\eta)}\overline{\ell}\ell.
\end{equation}
Meanwhile, Eq.(20) becomes
\begin{flalign}
\begin{split}
\mathscr{L}_{\ell}'=&+i\frac{g}{\sqrt{2}}\left[W_{\mu}^{+}\overline{({\nu}_{\ell})}_{L}\gamma_{\mu}{\ell}_{L}+W_{\mu}^{-}\overline{\ell}_{L}\gamma_{\mu}{({\nu}_{\ell})}_{L}\right]\\
&-\frac{i}{2}\sqrt{g^{2}+g'^{2}}Z_{\mu}\left[{\overline{({\nu}_{\ell})}_{L}\gamma_{\mu}({\nu}_{\ell})_{L}
-\overline{\ell}_{L}\gamma_{\mu}\ell_{L}}\right.\\&\left.+2\sin^{2}\theta_{W}\overline{\ell}\gamma_{\mu}{\ell}\right]-i\frac{gg'}{\sqrt{g^{2}+g'^{2}}}A_{\mu}\overline{\ell}\gamma_{\mu}{\ell}\\
&-{{\frac{g}{\sqrt{2}}\frac{M_{H}}{M_{W}}\sin(2\theta_{\ell})(\upsilon+\eta)}}\overline{\ell}\ell.
\end{split}
\end{flalign}
This result is completely consistent with the standard WS model.

From Eq.(22), we can obtain the masses of the charged leptons as
\begin{equation}
m_{\ell}=\sqrt{2}{M_{H}}\sin(2\theta_{\ell}).
\end{equation}
Eq.(24) shows that the difference of masses among the charged
leptons of three generations comes from the different left-right
mixing angles $\theta_{\ell}$ of their flavor eigenstates.

Comparing Eq.(11) with Eq.(24), the usual Yukawa coupling constant
$G_{\ell}$ is given by
\begin{equation}
G_{\ell}={\frac{\sqrt{2}}{\upsilon}M_{H}\sin(2\theta_{\ell})}={\frac{g}{\sqrt{2}}\frac{M_{H}}{M_{W}}\sin(2\theta_{\ell})}.
\end{equation}
From Eq.(24), we can obtain
\begin{equation}
\sin(2\theta_{\ell})=\frac{m_{\ell}}{\sqrt{2}M_{H}},
\end{equation}
where $m_{\ell}\ll \sqrt{2}M_{H}$ means $\theta_{\ell}\sim\
m_{\ell}/2\sqrt{2}M_{H}\ll1$. Now put $m_{e}\sim0.511MeV$,
$m_{\mu}\sim105.66MeV$, $m_{\tau}\sim1776.86MeV$ and
$M_{H}\sim125GeV$, one can obtain the magnitude of the three
left-right mixing angles are about
\begin{flalign}
\begin{split}
&\theta_{e}\sim1.45\times10^{-6},\\
&\theta_{\mu}\sim2.99\times10^{-4},\\
&\theta_{\tau}\sim5.03\times10^{-3}.
\end{split}
\end{flalign}
This result implies that the charged leptons, as the superposition
states of left- and right-handed elementary Weyl spinors, have an
extreme left-right asymmetry. It is noteworthy here that the
existence of left-right asymmetry should be related to the
destruction of parity conservation in weak interactions [10].

\section{Conclusions}

The present paper mainly discusses the origin of the masses of
charged leptons. The proposed model is based on two basic
hypotheses: the first one is that there exists a universal
constant of the Yukawa coupling term, this constant has the same
order of magnitude as the gauge coupling constant; the other one
is that the flavor eigenstates of the charged leptons are the
superposition states of left-handed and right-handed Weyl spinors
with different weights before electroweak symmetry breaking. Based
on these two hypotheses, it is found that the difference in masses
among the three generations of charged leptons is due to the
difference in the left-right mixing angles in the definition of
the flavor eigenstates of the different charged leptons. In fact,
the lines of thought presented in this paper may be applied to
three generations of neutrinos and even to the three generations
of quarks. The origin of neutrino masses is one of the most
compelling evidence for physics beyond the Standard Model (SM)
[11], and there is no reliable theory to explain it. We will
discuss the problem of neutrinos in our forthcoming works.

\end{document}